\documentclass[12pt]{JHEP3}

\newif\ifPDFLATEX
\PDFLATEXtrue

\usepackage{amsmath,epsfig}
\usepackage{amssymb,amsfonts}
\usepackage{latexsym}
\usepackage{epsfig}

\def\trule{\noalign{\hrule}}
\relax
\renewcommand{\theequation}{\arabic{section}.\arabic{equation}}
\def\be{\begin{equation}}
\def\ee{\end{equation}}
\def\bea{\begin{eqnarray}}
\def\eea{\end{eqnarray}}
\def\hri#1#2{\href{http://arxiv.org/abs/#1}{[ArXiv:#1]#2}}
\def\hre#1#2{\href{http://arxiv.org/abs/#1/#2}{[ArXiv:#1/#2]}}

\newcommand\fverb{\setbox\pippobox=\hbox\bgroup\verb}
\newcommand\fverbdo{\egroup\medskip\noindent%
                        \fbox{\unhbox\pippobox}\ }
\newcommand\fverbit{\egroup\item[\fbox{\unhbox\pippobox}]}

\newcommand{\bear}{\begin{eqnarray}}

\newcommand{\eear}{\end{eqnarray}}
\newbox\pippobox

\def\6{\partial}

\def\a{\alpha}

\def\e{\epsilon}
\def\m{\mu}
\def\n{\nu}

\def\sp{\;\;\;,\;\;\;}

\def\sq
\def\a{\alpha}
\def\b{\beta}

\def\cu{{\cal U}}
\def\cd{{\cal D}}
\def\ch{{\cal L}}

\large
\title{Discriminating MSSM families in (free-field) Gepner Orientifolds}

\author{{\large Elias Kiritsis$^{1,2}$, Bert Schellekens$^{3,4,5}$ and Mirian
Tsulaia$^{6}$}
\\
~\\
~\\
$^1$CPHT, Ecole Polytechnique, CNRS,
 91128, Palaiseau, France\\
~\\
$^2$Department of Physics, University of Crete,
71003 Heraklion, Greece\\
~\\
$^3$NIKHEF,
Kruislaan 409, 1009DB Amsterdam,
The Netherlands\\
~\\
$^4$IMAPP, Radboud Universiteit Nijmegen, The Netherlands\\
~\\
$^5$Instituto de F\'\i sica Fundamental, CSIC, Madrid, Spain\\
~\\
$^6$Institute for Theoretical Physics, Vienna University of
Technology, Wiedner Hauptstrasse 8-10, 1040, Vienna, Austria }

\preprint{
CPHT-RR059.0808 \\
NIKHEF/2008-012\\
TUW-08-08 \\ }      

\abstract{A complete analysis of
orientifold compactifications
involving Gepner models that are free fields (k=1,2) is performed.
A set of tadpole
solutions is found that are
variants of a single chiral spectrum. The vacua found have the property that
different families have different U(1)
charges so that one family cannot obtain masses in perturbation theory.
Its masses must come from instantons, allowing for a hierarchy of masses.
The phenomenological aspects of such vacua are analyzed. }

\begin{document}
\newpage
\section{Introduction and conclusions}

The search for vacua of
string theory that resemble the SM has a 24 year history, and is ongoing.
In the last decade, orientifold vacua
attracted a lot of attention in this respect as it became understood that
they allow a bottom-up approach   \cite{akt,akt1,aiqu}
in assembling the SM ingredients.
There are many distinct
ways of embedding the Standard Model
group into that of quiver
gauge theories, which appear in the context of orientifolds
and these are reviewed in \cite{d-review-1}-\cite{d-review-4}.
A general framework for
classifying such embeddings
in orientifolds, in particular that of the hypercharge,
was developed in \cite{adks} based on some mild assumptions.
This framework was applied
to orientifolds that can be
constructed from Gepner models (studied earlier in
\cite{Angelantonj:1996mw,Blumenhagen:1998tj,Blumenhagen:2004cg,Brunner:2004zd,Dijkstra:2004ym,Aldazabal:2004by,dhs}), using the algorithmic techniques of
RCFT developed in \cite{Fuchs:2000cm}.
A total of 19345 chirally distinct top-down spectra were found,
 that comprise so far the
most extensive such list known in string theory, \cite{adks}.
 For 1900 of these spectra at least one tadpole solution was also found.
 Combined with earlier
results for vacua realizing the
Madrid incarnation \cite{imr} of the Standard Model, \cite{dhs}
they contain the largest collection of
 vacua (tadpole solutions)  chirally realizing the (supersymmetric) SM.

Unfortunately, further progress in this
direction is hampered by the
fact that the tools to calculate the superpotential
and other important low energy quantities are not yet
so well developed.

In this paper we will focus on a
small subset of such vacua  that share a simplifying property:
 their CFTs and BCFTs can be constructed out of free fields.
 It is known that there
are two Gepner models
that are equivalent to free
fields. The $k=1$ model is equivalent to a free boson with $c=1$,
 \cite{wat,kir} while the $k=2$ model
is equivalent to a free boson and an Ising fermion with central charge $c=3/2$
\cite{Friedan:1989yz}.
 There are several ways of
tensoring these two models in order to
construct an orientifold compactification. We find that only vacua
 made out of six copies of the $k=2$ model
have the potential to produce spectra that resemble those of the SM.
 It is  such vacua that we will focus on this paper.
For the other tensor combinations of free field N=2 minimal models,
 namely (1,1,1,2,2,2,2), (1,1,1,1,1,1,2,2)
and (1,1,1,1,1,1,1,1,1) not even a SM configuration without
 tadpole cancellation ({\it i.e.}
the analog of a local model) was found in \cite{adks}.

 Our goal here is two-fold.
First to make a detailed
and extensive search
for orientifold vacua that are chirally similar to the supersymmetric
 SM\footnote{The search performed in \cite{adks} was not complete,
but it rather focused on finding the largest possible number of chirally
 distinct examples}.
 Second, to provide a qualitative
phenomenological study of the
tadpole solutions found, in order to
assess their potential to provide phenomenologically
 acceptable and interesting realizations of the SSM.
 If both of the above goals are
achieved successfully, the road is
open to a detailed calculation of the effective potential and interactions.

Our results are summarized as follows:

 $\bullet$  There are 96
tadpole solutions found in the ${(k=2)}^6$
compactification that all
realize the chiral spectrum No. 14062 in the classification of
  \cite{adks}. They give rise to 8
distinct massless spectra. There are
two possible hidden sector gauge groups: Sp(2) and O(2).
  The eight spectra differ apart from the hidden sector gauge group also in the
  non-chiral spectrum of massless particles.

$\bullet$ If we relax the
assumption made in \cite{adks}
that no chiral observable-hidden matter is present,
then we find three more
chirally distinct spectra, Nos. 101, 559, 800 in the list of \cite{adks}.
These include Pati-Salam
models but we will not study them further in this paper.

$\bullet$ Only the tadpole
solutions with a hidden Sp(2) group
have a phenomenologically sufficient number of right-handed neutrinos.

$\bullet$ There ~are ~three ~U(1) ~gauge
~symmetries, ~two ~of ~which ~are ~free ~of
~four-dimensional ~anomalies and one is ``anomalous".
 One of the two non-anomalous ones is hypercharge. The
other has a massive gauge boson and is
therefore expected to be violated by string-instanton effects.

$\bullet$ In order for this solution
to be phenomenologically viable, other points
in its moduli space must be chosen,
so that the massless non-chiral exotics obtain
sufficiently high masses in order to satisfy experimental constraints.

$\bullet$ One of the three families
has different charges under the
two ``anomalous" U(1) symmetries compared to the other two.
This has as a consequence that
selection rules for low energy
couplings are in effect. In particular, this family remains massless
 in perturbation theory.

$\bullet$ There is a single pair of Higgs multiplets

$\bullet$ A $\mu$-term is allowed and must therefore be tuned to small values.

$\bullet$ To protect low-energy lepton
number conservation discrete symmetries must operate.
Baryon number is violated only by SU(2)$_{\rm weak}$ instantons.

$\bullet$ The Fayet-Iliopoulos terms appearing
in the low-energy potential are shown to be zero at the
tadpole solution point. They must be kept zero as we move in moduli space.
As a byproduct we generalize to
arbitrary CFTs/BCFTs previous proofs on the vanishing
of loop corrections to the FI terms provided tadpoles cancel.

$\bullet$ String instanton corrections are
necessary (and are classified) in order for the third family to acquire masses.

$\bullet$ The expected pattern
of the neutrino mass matrix is of the
see-saw type allowing for light neutrino masses.

$\bullet$ Although the branes are not in a ``unified"
configuration, $\sin^2\theta_W=\frac{6}{13}$ at the string scale and
differs by less than 20\% from the unified value
of $\frac{3}{8}$. Therefore, a  change in
the masses of the charged non-chiral massive particles
 can  accommodate a conventional ``unification" of gauge couplings.

$\bullet$ The strong dynamics of
the hidden non-abelian gauge group
can trigger supersymmetry breaking.
However, to obtain an acceptable scale,
appropriate threshold
corrections must be advocated just below the string scale.

Although the results indicate that
this class of vacua are potentially
 compatible with phenomenology,
this requires also several special conditions to be met.
A lot of detailed analysis is necessary in
order to achieve this and we hope to report
on this in a subsequent publication.

\section{The tadpole solutions of the Gepner $(k=2)^6$ orientifold SM}

In this paper we consider the tensor
product of six $N=2$ minimal superconformal
field theories with $k=2$. The
central charge of each factor is $\frac32$, so that the internal CFT has
$c=9$, equivalent to six
free bosons and fermions.
Each $k=2$ factor has 24 primary fields. Each factor is
equivalent to the tensor
product of a free boson with 8 primaries and an Ising model. This means that
the resulting CFT can be
realized in terms of free fields, in contrast to most other $N=2$ minimal
model tensor products (a.k.a. Gepner models).
\def\Zbf{{\bf Z}}
However, in the construction of
modular invariant partition
functions (MIPFs), orientifolds and tadpole solutions
no use is made of the specific free
field theory properties of these models. After tensoring the six factors with
the space-time NSR fermions,
imposing world-sheet supersymmetry by extending the chiral algebra with
the product of all
world-sheet supersymmetry generators,
and extending the chiral algebra to obtain
space-time supersymmetry, we end up with a
CFT with 2944 primary fields, 512 of which are simple currents.
Under fusion, these simple currents
close to form a
discrete group $\Zbf_4 \times \Zbf_4 \times \Zbf_4 \times \Zbf_2$.

We now build all the MIPFs that
can be constructed using these simple currents, using the
algorithm of \cite{Gato-Rivera:1991ru}\cite{Kreuzer:1993tf}.
In normal circumstances all these MIPFs would be distinct, but in
this case there are two special
circumstances: a permutation symmetry among the six identical factors,
and the fact that each factor
contains an Ising model. A special feature of the Ising model is that
its simple current MIPF is
identical to the diagonal invariant. This happens because the only simple
current orbit with
charge $\frac12$ happens to be a
fixed point of the simple current (this orbit is formed by the
spin field of the Ising model).
This degeneracy extends to products of Ising models, and as a result some
generically distinct MIPFs are actually identical.

The permutation symmetry occurs frequently
in other Gepner models, and we deal with it by
considering only one
member of a permutation orbit. The Ising degeneracy occurs only in  a few
cases and can be dealt with
by comparing the resulting MIPFs. The only problem is that there
is some interference between
the two degeneracies. It may happen that an Ising degeneracy does
not occur between the selected
representatives of the permutation orbits, but between other members.
This will then result in some over-counting.

Although this degeneracy can be
removed in principle\rlap,\footnote{This would involve acting with all 720
permutations on all MIPFs, but this is not
completely straightforward. First one has to work
out how the permutation acts
on resolved fixed points, {\it i.e.} distinct fields that come from the
same combinations of minimal model primaries.}
we have not implemented this because the overcounting is
only a minor problem. After removing
permutations and identical MIPFs we end up with 1032 MIPFs, and
we expect the actual number of
distinct ones to be slightly smaller than this. For each MIPF we construct
all simple current orientifolds,
according to the
prescription of \cite{Fuchs:2000cm}. The total number of distinct
orientifolds (taking into account
known orientifold equivalences as described in \cite{Fuchs:2000cm} and the
permutation symmetry) ranges
from 4 to 64, depending on the MIPF. This includes some zero-tension
orientifolds that are of no
further interest, since the dilaton
tadpole forbids all Chan-Paton multiplicities.

For each MIPF we then compute all
boundary states, using the formula given in \cite{Fuchs:2000cm}.
To each of these cases
we then search for standard model configurations. Here we
apply the same search algorithm used already in  \cite{adks}
for the other Gepner models. The only
difference is that we remove the upper limit on the
number of boundary states, which
was set at 1750 in \cite{adks} for purely practical reasons. In the
case of the $2^6$ model,
only a handful of MIPFs exceed that limit, and therefore we decided to
do a complete scan. This did not yield anything new, though.
Indeed, the standard model configurations
we describe below were all
already found during the search performed in \cite{adks}.

The last step in the procedure is to
try and solve the tadpole conditions for the hidden sector, in order
to cancel all tadpoles
introduced by the orientifold and
the Standard Model configuration. Here too we went
slightly beyond \cite{adks} by allowing
chiral matter between the observable and the hidden sector. Normally
this produces such a huge number of
solutions that it is  preferable to require observable/hidden matter
to be non-chiral. While chiral
observable/hidden matter is not necessarily a phenomenological disaster
(and can even be desirable in certain circumstances),
it does require additional mechanisms to make
it acquire a mass. In this particular case,
however, we already knew that the number of tadpole solutions
was extremely small, so it seemed worthwhile to try and relax the criteria.

The search of \cite{adks} produced a total
of about 19000 chirally distinct standard model
configurations, and tadpole solutions
were found for 1900 of them. In the new search for the
$2^6$ model we found tadpole
solutions for 4 models. On the list of 19000 (ordered according to
the first time each spectrum occurred\rlap,\footnote{Note that
in \cite{adks} they were
ordered according to frequency} and
available on request) these were nrs. 101, 559, 800 and
14062. Only in the latter
case did we find solutions with non-chiral observable/hidden matter. This
means that this last case was
within the scope of \cite{adks}. Nevertheless, the tadpole solutions
were not found at that
time for a very simple reason: no attempt was made to solve the
tadpole conditions for a
certain model if a solution was already known. In this particular case,
there turns out to exist a
solution for spectrum nr. 14062 for Gepner model (2,2,2,6,6), which was
found first. It was presented in \cite{adks}
in section 6.5, as a ``curiosity". This model is rather similar
to the ones presented here,
but the (2,2,2,6,6) model is not a
free CFT. It is in fact so similar (including
non-chiral matter, which is
not taken into account when comparing spectra) that we expect that
these models are actually related,
presumably by an orbifold procedure that maps three copies
of $k=2$ to two copies of $k=6$, but we have not investigated this.

In all orientifolds of all MIPFs of the
tensor product $2^6$ the spectrum 14062 occurred 168 times, and
in 96 cases there was a
solution to the hidden sector tadpole equations. These solutions occurred for
the following MIPF numbers:
41, 414, 415, 416, 417, 418, 644, 646, 651, 652, 662, 1018, 1021. These
numbers are labels assigned by
the generating program ``kac" to the 1032 MIPFs, and are listed
here in order to identify the MIPFs and reproduce them, if necessary.

For comparison we give here the total number of boundary state configurations
with at least one tadpole solution for the other models: 43008 for
nr. 800, 168 for nr. 559 and 6144 for nr. 10.
Note that this is not the {\it total}
number of tadpole solutions: any given boundary state configuration
may admit many, often a huge number,
of tadpole solutions. We only explored the full set
of solutions for spectra of type 14062.
As already mentioned above, all
tadpole solution for spectrum types 800, 559 and 101 contain
chiral observable-hidden matter.
For these three configurations there are no tadpole solutions
without such chiral exotics.
On the other hand, for spectrum 14062
all tadpole solutions are free
of chiral exotics. In fact, in some cases there
is no observable-hidden matter at all.

Spectrum 14062 has a Chan-Paton group $U(3) \times Sp(2) \times U(1) \times U(1)$, with the
hypercharge realized as in the familiar ``Madrid" configuration \cite{imr}, but with an interchange of the
r\^oles of brane c and d for some of the quarks and leptons. In contrast to the Madrid models, which
with very rare exceptions have an exact $B-L$ gauge symmetry, all superfluous $U(1)$'s in these
models are broken, so that the surviving gauge symmetry (apart from the hidden sector) is
exactly $SU(3) \times SU(2) \times U(1)$. We will discuss these spectra in much more detail in the
next section. Usually we will denote the Chan-Paton factor $Sp(2)$ as $SU(2)$ when its orientifold
origins are unimportant.

\begin{table}
\begin{center}
\begin{tabular}[t]{|c|c|c|c|c|c|c|c|}\trule
MIPF id & Order & $h_{11}$  & $h_{12}$ & Singlets  & Sol. Types & Total & Glob. An? \\ \trule
41           & 32      & 11 & 17 & 223 & 1112 & 4 &  No \\ \trule
414           & 32      & 17 & 11 & 223 & 3345 & 4 &  No \\ \trule
415           & 32      & 11 & 17 & 223 & 1112 & 4 &  No  \\ \trule
416           & 32      & 17 & 11 & 223 & 3345 & 4 & Yes  \\ \trule
417           & 32      & 9 & 15 & 219 & 6678 & 16 & Yes \\ \trule
418           & 32      & 11 & 17 & 223 & 1112 & 4 & Yes  \\ \trule
644           & 128      & 11 & 17 & 223 & 1112 & 4 & No  \\ \trule
646           & 128      & 9 & 15 & 219  & 6678 & 12 & Yes  \\ \trule
651           & 128      & 17 & 11 & 223 & 3345 & 4 & No \\ \trule
652           & 128      & 17 & 11 & 223 & 3345 & 4 & Yes \\ \trule
662           & 128      & 9 & 15 & 219 & 6678 & 12 & Yes   \\ \trule
1018           & 32      & 9 & 15 & 219 & 6678 & 8 & Yes \\ \trule
1021           & 32      & 9 & 15 & 219 & 6678 & 16 &  No \\ \trule
\end{tabular}
\end{center}
\caption{The MIPFs with tadpole solutions}
\label{tabMIPF}
\end{table}

Table (\ref{tabMIPF}) lists the main characterizations of the MIPFs for which
tadpole solutions for spectrum 14062 exist.  We specify
the order of the simple current subgroup that produces them, the Hodge numbers
of the compactification and the
number of singlets in the spectrum for the corresponding
heterotic string theory.
The number of boundary states is 320 in all cases, and the gauge
group in the heterotic theory is $E_6 \times E_8 \times U(1)^5$ in
all cases. Of course the
orientifolds we
construct are based
on a type-IIB theory, and heterotic data are only given
here as a way to characterize the MIPF.

The simple current group
is $\Zbf_4 \times \Zbf_4 \times \Zbf_2$ if the order is 32 and
$\Zbf_4 \times \Zbf_4 \times \Zbf_2\times \Zbf_2\times \Zbf_2$
if the order is 128. Note that
the list of Hodge numbers is not mirror symmetric.
The complete list of Hodge numbers of
the $2^6$ tensor product {\it is}
mirror symmetric, even if one includes the number of singlets
and gauge bosons. However,
mirror symmetry does not extend to the boundary states, indeed not
even to the total number of boundary states.
Nevertheless, there do exist MIPFs with Hodge data (15,9,219) and
even precisely 320 boundary states, but they did not produce any solutions.

Columns 6 and 7 specify some
information concerning the tadpole solutions we found.
In column 7  we indicate
for how many standard model configurations at least one solution
exists. It turns out that in
each of those
cases ({\it i.e.} 96 in total) there are in fact four solutions
to the tadpole conditions,
one with a hidden sector gauge group $Sp(2)$, and three with an
$O(2)$ hidden sector group.
Of the total number of $4\times 96 = 384$ solutions only 8 are
different.
 In column 6 we indicate which of those eight solutions
occur for each MIPF.
This turns out to depend only on the MIPF, and not on the standard model
configuration.
Note that the kind of solution
that occurs correlates perfectly with the Hodge data.

\begin{table}
\begin{center}
\begin{tabular}[t]{|c|c|c|c|c|c|c|c|c|}\trule
Spectrum & H & $Y_A$ & $Y_S$ & $P_A$ & $P_S$ & $R$ & $T$ & $X$ \\ \trule
1 & $Sp(2)$ & 0 & 4 & 0 & 2 & 2 & 1 & 4 \\ \trule
2 & $O(2)$ & 0 & 4 & 0 & 2 & 3 & 0 & 0 \\ \trule
3 & $O(2)$ & 4 & 0 & 0 & 2 & 1 & 2 & 4 \\ \trule
4 & $Sp(2)$ & 4 & 0 & 0 & 2 & 0 & 3 & 4 \\ \trule
5 & $O(2)$ & 4 & 0 & 0 & 2 & 1 & 2 & 0 \\ \trule
6 & $O(2)$ & 2 & 2 & 2 & 0 & 1 & 2 & 4 \\ \trule
7 & $Sp(2)$ & 2 & 2 & 2 & 0 & 0 & 3 & 4 \\ \trule
8 & $O(2)$ & 2 & 2 & 2 & 0 & 1 & 2 & 0 \\ \trule
(2,2,2,6,6) & $U(2)$ & 4 & 0 & 0 & 2 & 0 & 0 & 0 \\ \trule
\end{tabular}
\end{center}
\caption{The distinct spectra and their non-chiral exotics. The first eight occur in the
$(2,2,2,2,2,2)$ tensor product and are the subject of this paper. The last one has been found
in \cite{adks} for the (2,2,2,6,6) tensor product.}
\label{tabSpectra}
\end{table}

The eight distinct spectra
are tabulated in table  (\ref{tabSpectra}). All eight spectra have identical
chiral states, which
we specify in the next section.
Here we just focus on the differences, which consist of
the choice of hidden sector gauge groups, and some non-chiral exotics.
Column two lists
the hidden gauge group $H$. The other columns specify the multiplicities of
the seven kinds of non-chiral
exotics that may occur. We have named them $Y_A$ $\ldots$ $X$, and
in table (\ref{tabEx}) we indicate their Chan-Paton representations.  For comparison
we have also listed the (2,2,2,6,6) model presented in  \cite{adks} in table (\ref{tabSpectra}).  It has
an $U(2)$ hidden sector group with the rare feature of being completely hidden, by not having any massless
matter at all (of course there do exist massive excited states in all open string sectors). Note also that all
these spectra, including the last,
have the same total number of non-chiral rank-2 exotics for each of the a,b,c and d branes, which
may be distributed in different ways over symmetric and anti-symmetric representations.

An important additional constraint
is the absence of global anomalies. In RCFT models, this
leads to a large
number of necessary conditions obtained by adding probe branes to a given
model, as discussed in  \cite{Uranga:2000xp}.
Since the probe branes at our disposal are limited by ``rationality"
of the RCFT,
it is not guaranteed
that this exhausts all possible origins of global anomalies, but
we do take into account all the ones we can.
In Gepner orientifolds these constraints
eliminate some models, but
their effect is limited to
rather few tensor combinations, and is not extremely
restrictive even in those
cases \cite{Gato-Rivera:2005qd}.
Also in the present class there turn out to be
tadpole solutions with
global anomalies, but they were already eliminated from the set
discussed above. In column 8 of table 1 we
indicate in which cases there were additional tadpole solutions
with global anomalies. Note
that these anomalous solutions do not correlate with the Hodge data.

\begin{table}
\begin{center}
\begin{tabular}[t]{|c|c|c|c|c|c|c|}\trule
$U(3)_a$ & $SU(2)_b$ & $U(1)_c$  & $U(1)_d$ & $H$ & Y  & Symbol \\ \trule
A           & 0     & 0 & 0 & 0 & $\pm\frac13$ &  $Y_A$ \\ \trule
S           & 0     & 0 & 0 & 0 & $\pm\frac13$ &  $Y_S$ \\ \trule
0           & 0     & 0 & A & 0 & $ 0 $ &  $P_A$ \\ \trule
0           & 0     & 0 & S & 0 & $\pm 1$ &  $P_S$ \\ \trule
0           & 0     & 0 & 0 & A & $ 0 $ &  $R$ \\ \trule
0           & 0     & 0 & 0 & S & $ 0 $ &  $T$ \\ \trule
0           & 0     & V & 0 & V & $ \pm\frac12 $ &  $X$ \\ \trule
\end{tabular}
\end{center}
\caption{The non-chiral exotics that may occur in the eight distinct models.}
\label{tabEx}
\end{table}

\section{The low-energy characteristics of  tadpole solution No. 1}

All of the tadpole solutions we presented in the previous section are missing
 two right-handed singlets in the SM stack.
 Overall SM singlets, even if they do
not come from the SM stack can in
principle play the role of right-handed neutrinos.
 A look at table \ref{tabSpectra}
shows that  global singlets with
zero mass are the multiplets labeled $R$ for the hidden $Sp(2)$
 group\footnote{There is another
interesting possibility: that we choose as such singlets
  the fist string level descendants of the T multiplets.
 As the projection alternates
between the string levels these will be global singlets.
 In this case in the spectrum No. 1 of table \ref{tabSpectra}
 we may consider an extra
three right-handed neutrino singlets, two of the R type and one
 of the T type. In spectra Nos. 4 and 7, all such neutrino singlets are of
 type T} or the multiplets $T$ for the hidden $SO(2)$.

It is preferable for phenomenological
 reasons (supersymmetry breaking in particular)
to a have a strongly-coupled gauge group in the hidden sector.
The presence of a sufficient number of
 right-handed neutrinos\footnote{In cases where large internal
volume is present, even a smaller number of right-handed
 neutrinos can be phenomenologically acceptable.
This works via the presence and
mixing of suitably light KK states and as shown in \cite{akt1}
 it is not far from the current data of
the neutrino sector. Finally, even in the complete absence of right-handed neutrino candidates,
 neutrino masses and mixings can be generated by higher dimension operators mediated by instantons \cite{stringinst}. } and the requirement of a non-abelian hidden sector
  therefore selects spectrum No. 1, which has a hidden $Sp(2)$ group. The complete
  spectrum of this solution is shown in table \ref{tab1}.

The solutions we find have
unbroken N=1 supersymmetry in four
dimensions, therefore each entry of  table \ref{tab1}
 corresponds to an N=1 chiral multiplet.
The N=1 vector multiplets for all gauge groups are assumed.
As usual $V$ stands for the vector representation,
$V^*$ for the conjugate vector representation,
 $S$ for the two-index symmetric representation while
$A$ stands for the two-index
antisymmetric representation. In particular for a U(1) gauge group,
 $V$ indicates charge $+1$, $V^*\to -1$, $S\to +2$, while $A$ indicates
a missing massless
particle (although the associated stringy tower is intact as the
 projection alternates at
alternate string levels). Dimension gives the total number of multiplets
 independent of chirality, while Chirality gives the net
 chiral number of multiplets.
Chirality is + by
convention for left-handed
fermions and its minus for
left-handed fermions.
Dimension=3, Chirality=3
therefore means that there are 3 left-handed multiplets.
while dimension=3,
chirality=-1 means there are 2 right-handed and one left-handed multiplets.

The hypercharge tabulated in table \ref{tab1} is given by
\be
Y={1\over 6}Q_3-{1\over 2}Q_c-{1\over 2}Q_{d}
\label{1}\ee
whose gauge boson is
massless in this solution\footnote{As observed in \cite{dhs,adks}
this condition seems to be the strongest constraint
towards finding a SM-like vacuum  in Gepner orientifolds.}.
This is the Madrid hypercharge
embedding or $x={1\over 2}$ in the global classification of \cite{adks}.

\begin{table}
 \begin{tabular}[t]{|c|c|c|c|c|c|c|c|c|}\trule
 Dim & Chirality & U(3)$_a$ & SU(2)$_b$ & U(1)$_c$ & U(1)$_d$&SU(2)$_h$& Y& Symbol \\ \trule
 3&3 &V &V &0 &0 & 0& \phantom{m}$+{1\over 6}$\phantom{m} &  Q \\
   2& -2& V& 0& V$^*$&0 &0 &\phantom{m}$+{2\over 3}$\phantom{m}  &$ U$ \\
    2& -2& V& 0& V&0 &0 & \phantom{m}$-{1\over 3}$\phantom{m} &$D$ \\
      3& -1& V& 0& 0& V$^*$& 0& \phantom{m}$\pm{2\over 3}$\phantom{m} &${\cal U}$ \\
       1& -1& V& 0& 0& V&0 & \phantom{m}$-{1\over 3}$\phantom{m}&$ {\cal D}$ \\
     2& 2&0 & V& 0& V& 0& \phantom{m}$-{1\over 2}$\phantom{m}&L\\
       3& 1& 0& V& V& 0& 0& \phantom{m}$\pm{1\over 2}$\phantom{m}&K\\
        3& -3& 0& 0&V & V&0 &  \phantom{m}$-{1}$\phantom{m}&$ E_R$ \\
          1& 1& 0& 0& V& V$^*$&0 &  \phantom{m+}${0}$\phantom{m}&$N_R$ \\
\trule
              4& 0&S & 0& 0& 0& 0& \phantom{m}$\pm{1\over 3}$\phantom{m}&$Y_S$ \\
 2& 0& 0& 0& 0& S& 0& \phantom{m}$\pm{1}$\phantom{m}&$P_S$ \\
            4& 0& 0& 0&V &0 & V& \phantom{m}$\pm{1\over 2}$\phantom{m}&$X$   \\
\trule
   2&0 & 0& 0& 0& 0& A& \phantom{m+}${0}$\phantom{m} &$R$ \\
    1& 0& 0&0 &0 &0 &S &  \phantom{m+}${0}$\phantom{m}&$T$ \\
\trule
 \end{tabular}
\caption{The massless spectrum of tadpole solution No. 1 of spectrum 14062.}
\label{tab1}
\end{table}


  The following massless states are charged (non-chiral) exotics beyond the MSSM:
  \begin{itemize}

  \item A pair of the up-like anti-quarks $\overline{\cal U}$.

  \item The 2 right-handed and 2
left handed $\bf 6$ representations of SU(3),
labelled $Y_s$ in table \ref{tab1}.
  Although they have
fractional hypercharge, all
colour singlets that one can make using them have integer electric charge.

  \item The 2 right-handed and 2
left handed multiplets labelled $X$ in table \ref{tab1}.
  They are doublets of the hidden $SU(2)_h$,
and have half-integer Y and electric charge.

   \item The 1 right-handed and 1
left handed multiplet labelled $P_s$ in table \ref{tab1}.
  They have charge $\pm 2$ under  $U(1)_d$
and have integer Y and electric charge.

  \item The one real
multiplet labelled $T$ in
table \ref{tab1} transforming as the adjoint of the hidden $SU(2)_h$ group.

    \end{itemize}

Finally we should stress that the two
chiral multiplets
labeled $R$ in table \ref{tab1}
are absolute singlets (as the antisymmetric of
$SU(2)_h$ is a singlet)
and are expected to play the role of the missing 2 right-handed neutrinos.

Because of the above fields the
particular point in the moduli space where the tadpoles were solved
 is not suitable for describing the low-energy world.
It is natural to assume that by moving a distance
of order of the string scale in
moduli space such non-chiral states will acquire masses
which may be anywhere
from 100 TeV to the string scale so they are directly unobservable.
 Of course such particles may have indirect effects in the low energy physics.
Below we will consider
all possible non-renormalizable
superpotential terms and therefore we are sure to
include all indirect effects due to these massive states.

Therefore in the sequel we
will assume that the
multiplets $Y_s$, $X$, $P_s$,
 $T$ and one non-chiral pair of the $\overline{\cal U}$ quarks
are massive and have been integrated out.

\subsection{Anomalies}
It is by now well known that generic U(1) gauge symmetries in
 orientifold vacua are anomalous.
Their anomalies are canceled by the GS mechanism that
in four dimensions
involves closed string axion scalars \cite{iru}.
In the process,  the associated gauge bosons acquire a mass that is
 generically moduli dependent \cite{akr,ana} and the gauge symmetry is broken.
Unless the associated global symmetry is also spontaneously broken by
D-terms, it survives in
perturbation theory and is only broken by gauge instantons.

It is important to stress that a U(1) gauge symmetry can be broken and its
associated gauge boson
acquires a mass even when the U(1) in question has no four-dimensional
anomalies. This phenomenon
was observed in \cite{imr,akr} and was explained in \cite{ana}.

In the vacuum at hand we can calculate the four-dimensional
 mixed anomalies of the
three U(1) factors. The results are in table \ref{tab3}.

\begin{table}
\begin{center}
\begin{tabular}[t]{|c|c|c|c|}\trule
  & U(1)$_3$ & U(1)$_c$ & U(1)$_d$ \\ \trule
SU(3)$_a$   &0  & 0 & 0 \\ \trule
SU(2)$_b$     &9& 1 &  2\\ \trule
SU(2)$_h$     &0& 0 &  0\\ \trule
gravity  & 0 & 0 & 0 \\ \trule
\end{tabular}
\end{center}
\caption{The mixed four-dimensional anomalies of U(1)'s}
\label{tab3}
\end{table}

The anomaly  matrix is defined $K_{IJ}= Tr[Q_J(T^aT^a)_I]$,
where $J=3,c,d$, and $I=1$ corresponds to the colour $SU(3)$, $I=2$
corresponds to the
weak $SU(2)$, and $I=3$ corresponds to the mixed gravitational
anomaly $Tr{Q_J}$.

\begin{itemize}

\item Note that the only non-trivial
non-abelian anomaly is that with $SU(2)_b$. This implies that
there are two independent
U(1) combinations that are free from four-dimensional anomalies.
 We find however that
only one of them, the hypercharge in (\ref{1}) is massless.
 Therefore, all other U(1)'s except Y are massive.

\item $U(1)_a$ is baryon
number and it is violated only by $SU(2)_b$ instantons.
This violation is tiny and therefore baryon number is a very good global
 symmetry of this vacuum \cite{iq}.

 \item None of the two U(1)'s that are anomaly
free in four
dimensions ($aQ_a+cQ_c+dQ_d$ with $9a+c+2d=0$)
is violated by gauge instantons.
 Y remains massless and we
expect no violation due to
instantons.
 However we expect that
the other
anomaly-free U(1) symmetry it is
  broken by stringy instantons.

\item Although there are
anomalous U(1)'s and
mixed anomalies, the gravitational mixed anomaly is zero.

  \end{itemize}

  This vacuum has two extra
anomalous U(1)s beyond the SM symmetries. Extra (anomalous) U(1) symmetries
  are a generic prediction
of orientifold vacua,
their number ranging from a minimum of one to several, \cite{akt,ak}.
  The masses of such
gauge bosons can be low when
the string scale is low.
They can also be accidentally low even if the string scale is large
  in the case of highly
asymmetric compactifications, \cite{akr}.
The phenomenological consequences of anomalous U(1) gauge bosons in such cases
  have been explored in \cite{ak,giiq,cik}.
 A review on Z's from string theory can be found in \cite{Lang}.

\section{The low energy MSSM fields}

After integrating out the non-chiral
exotics we are left with
fields that are in one to one correspondence with the MSSM.
\footnote{Our conventions are the $I,I,K=1,2,3$, $i,j,k=1,2$.}
We have 3 quarks $Q^I$, two
up and down anti-quarks $U^i$, $D^i$,
of the first type, one
anti-quark of the second type: ${\cal U}$,
 one down anti-quark of the second type
${\cal D}$, two lepton
doublets, $L^i$, two left handed
lepton doublets $K^i$ that together
with the right-handed
doublet $\overline H$ will provide the third lepton double
 and the pair of MSSM Higgs,
three right-handed electrons $E^I$ and three (neutrino)  singlet $N$ and $R^i$.
They are all
summarized in table \ref{tab2} along with their various U(1) charges.

\begin{table}
\begin{center}
\begin{tabular}[t]{|c|c|c|c|c|c|c|}\trule

 Number & U(1)$_3$& SU(2)$_b$ & U(1)$_c$ & U(1)$_d$& Y &Chiral field \\ \trule


 3 &1 &{\bf 2}&0 &0 & \phantom{m}$+{1\over 6}$\phantom{m}&  Q$^I$ \\

   2& -1&{\bf 1}& -1&0&\phantom{m}$-{2\over 3}$\phantom{m}& $U^i$ \\

 2& -1&{\bf 1}& 1&0&\phantom{m}$+{1\over 3}$\phantom{m}   & $D^i$ \\

      1&  -1&{\bf 1}& 0& -1& \phantom{m}$-{2\over 3}$\phantom{m} & $ {\cal U}$ \\

       1& -1&{\bf 1}& 0& 1& \phantom{m}$+{1\over 3}$\phantom{m} & ${\cal D}$ \\

\trule

         2&0&{\bf 2} & 0& 1& \phantom{m}$-{1\over 2}$\phantom{m} &$L^i$\\

        2& 0&{\bf 2}& 1& 0& \phantom{m}$-{1\over 2}$\phantom{m} &  $K^i$\\

         1& 0&${\overline{\bf 2}}$& -1& 0& \phantom{m}$+{1\over 2}$\phantom{m}& $\overline H$\\

         3 &0&{\bf 1} &-1 &-1  & \phantom{m}$+1$\phantom{m}&$E^I$ \\

\trule

          1& 0&{\bf 1}& 1& -1 &\phantom{m} 0\phantom{m}& $N$ \\

          2& 0&{\bf 1}& 0& 0 &\phantom{m} 0\phantom{m}& $R^i$ \\\trule
          \end{tabular}
          \end{center}
\caption{The low energy MSSM states as left-handed chiral multiplets}
\label{tab2}
\end{table}

There are two immediate observations.
A $\m$-term $K^i{\overline H}$ is
not forbidden by the gauge symmetry in the superpotential
but we are at a
special point where this term is zero. There are two possibilities:
 (a) either this term
is forbidden by one of the discrete symmetries of the vacuum or
(b) this term is moduli dependent, and we happen to be at one of its zeros.
 In any case we will assume
that we are in a region of moduli space that this term is small
 compared to the string
scale and close to what is required for electro-weak physics.

The second observation
is that because one of the
lepton doublets (orthogonal to the one that mixes with ${\overline H}$)
has exactly the same quantum numbers,
(including the anomalous U(1) charges)
as the Higgs, we expect the lepton number to break at the renormalizable level.
In this theory, baryon number as we
will discuss later is
expected to be a very
good global symmetry as only SU(2) gauge instantons break it.\footnote{
There is also the
possibility that string
instantons break it, but we will not further
entertain  this possibility here.}
Because of this the
constraints on lepton
number violation are weak, but exclude however renormalizable couplings.
To proceed we will now write
 all quadratic and cubic terms in
the superpotential that are allowed by the gauge symmetries both anomalous and
non-anomalous.

The most general gauge-invariant quadratic superpotential is
\be \label{W2}
W_2=
~K\bar H +RR
\ee
 while the cubic one is
 \be
  \label{W3}
   W_3=Q U
K+Q D \bar H+Q{\cal U} L +L N{\bar H} +L E K +K\bar H R+RRR
\ee
where we have dropped
both the indices and coefficients
as we are interested in the qualitative features.

The following observations are relevant

\begin{itemize}

\item A linear term in $R$ is allowed in the superpotential as $R$ is
 a global singlet. This term is zero in the Gepner point, but may appear
in other regions of moduli space and along with gaugino condensation may
 trigger supersymmetry breaking.

\item It is reasonable to assume that
the role of Higgses is taken over by $\bar H$
 and a linear
combination of $K^i$.\footnote{There is the further possibility that
  $L^i$ also participate in electro-weak symmetry breaking.
   In that case the $\cu$ quark has a tree-level Yukawa coupling.}

\item The anti-quarks $\cu,\cd$ have no Yukawa coupling in $W_3$.

\item Because the Higgs $H$ and one
of the leptons have the same global quantum numbers, several couplings
violate lepton number.

\end{itemize}

Looking further we may
write down the most
general quartic superpotential consistent with the gauge symmetry,
\be
W_4=(QU)(LN)+(QD)(LE)+(QU)(QD)+(Q{\cal U})(Q{\cal D})+(LL)(EN)+K\bar H K\bar H+
\label{W4}\ee
$$
+(Q{\cal D})\bar H N+(Q{\cal D})EK+K\bar H RR+W_3 ~R$$

We observe that
if the right-handed
neutrino $N$ obtains a vev, the $\cd$ quark (not to be confused with the two quarks $D$) acquires a
Yukawa coupling, which will be very small for any acceptable value of the vev of $N$.

\section{Lepton number violation and discrete symmetries}

To avoid lepton number
violation at the observable level a discrete symmetry must be invoked.
This discrete symmetry must distinguish between
the two chiral
doublets $K^1,K^2$ that will provide one Higgs and one lepton
doublet. There may be several
such discrete symmetries but
the one that will do the job is the following $Z_2$ symmetry
\be
K^1\leftrightarrow K^2\sp L^i\to -L^i\sp E^I\to -E^I\sp
 N\to -N \sp R^i\to -R^i
\ee
If we now label $K^1+K^2\to H$ which
will now be the Higgs
and $K^1-K^2\to \ch$ which will now be the third lepton doublet,
we may rewrite the superpotentials that are invariant under such a symmetry
\be
\label{w2}
W_2=
~H\bar H +RR
\ee
 \be
  \label{w3}
   W_3=Q U
H+Q D \bar H+L N{\bar H} +L E H +\ch R\bar H
\ee
\be
W_4=(QU)LN+(QD)LE+(QU)(QD)+(Q{\cal U})(Q{\cal D})+LLEN+
\label{w4}\ee
$$
+\ch\ch \bar H\bar H+
HH\bar H\bar H+(Q{\cal D})E\ch +H\bar H RR+(QU)\ch R+(Q\cu)LR+
$$
$$
+LE\ch R+H\bar HRR+RRRR
$$

We observe that

\begin{itemize}

\item Lepton number is
preserved at the renormalizable level.
If the string scale and the scale of massive exotics is beyond
10 TeV or so,
this will also make
the non-renormalizable contributions to lepton number violation unobservable

\item  The $\cu$, $\cd$ quarks as
well as the electron singlet associated with $\ch$ remain massless.

\end{itemize}

Products of Gepner models
typically have large discrete symmetries.
These might be broken by the simple-current extensions procedure,
 as well as turning on closed string moduli.
It is however expected
that in subspaces of the
moduli space there are remnants of the discrete symmetry.
 As the previous analysis shows,
such symmetries are crucial
for the phenomenological viability of
this class of vacua and their presence must
 be carefully analyzed but this is beyond the scope of the present paper.

\section{The D-terms}

The general form of the $D$--term potential is
\begin{equation}
V_D = \sum_i D_i^2,
\label{2}\end{equation}
For the U(1)'s the D-term has the general form
\begin{equation}
D_i=\xi_i + \sum (q_i |X_i|^2)
\label{3}\end{equation}
where $q_i$ is the charge of the chiral superfield $X_i$ under the
corresponding gauge group $U(1)_i$, and $\xi_i$ is the associated FI term.
For the three U(1)'s of the model we have
\begin{equation}
D_a =\xi_a+ Q^I Q^{I \dagger} -
U^iU^{i \dagger} - D^i D^{i \dagger} - {\cal U}^i{\cal
U}^{i \dagger} - {\cal D}{\cal D}^\dagger ,
\label{4}\end{equation}
\begin{equation}
D_c =\xi_c - E^I E^{I \dagger}  - U^iU^{i \dagger} + D^i D^{i
\dagger} + H H^{ \dagger}+ \ch \ch^\dagger - {\overline H} {\overline H}^\dagger +
N N^\dagger ,
\label{5}\end{equation}
\begin{equation}
D_d =\xi_d - E^I E^{I \dagger} + L^i L^{i \dagger}-
{\cal U}^i{\cal U}^{i \dagger} +{\cal D}{\cal D}^\dagger - N
N^\dagger
\label{6}\end{equation}

{}The contribution from non - abelian D terms to the Higgs potential
 has the standard from
\begin{equation}
D^2_{SU(2)}= \frac{g^2 }{8}{(H {H}^\dagger
- {\overline H} {\overline H}^\dagger)}^2
+\frac{g^2}{2}(H {\overline H}^\dagger)({\overline H} H^\dagger)
\label{7}\end{equation}

Finally the $D$ -- term potential is
\begin{equation}
V_D = D_{a}^2+D_c^2+D_d^2+D_{SU(2)}^2
\label{8}\end{equation}

\subsection{The Fayet-Iliopoulos terms}

An important ingredient for the
phenomenology of orbifold models is the presence and size of FI terms.
FI terms can appear at disk level,
and their presence is typically tracked by a spontaneous breaking of the
associated U(1) global symmetry due
to the D-term potential they generate. An important question is whether a FI
term can appear at one loop if it is zero at tree level.
This was answered in the
negative in \cite{pop} where a calculation of the FI term was performed
 in the $Z_3$ orientifold, and was argued to hold for
more general orbifolds.
This was confirmed in the case of intersection $D_6$ branes in a flat background, \cite{aker}.
However it is not obvious
that such a conclusion
holds more generally for the RCFT vacua that we study here.

Consider a general orientifold
ground state based on an
arbitrary CFT and its BCFT. We assume that the CFT and BCFT realize a
ground state with N=1 spacetime
four-dimensional supersymmetry. Moreover, all
consistency conditions are satisfied at tree-level
(sphere and disk) and the
disk tadpoles have been canceled.
 All such assumptions are valid in the vacua we are considering made out of
RCFTs including  Gepner models.

Consider the U(1) gauge groups in
this ground state that may be anomalous,
but are massless at tree
level (the mass developed by anomalous U(1)'s
is a annulus
effect \cite{akr}.) This by
definition implies that their associated FI term is zero at disk order
as it would otherwise break the gauge symmetry or supersymmetry at tree level.
We will now show that no FI  term can be generated at one loop.

To track a non-zero FI term at one
loop we may calculate
the one-loop mass term of scalars charged under the U(1) in question.
Such scalars were  massless at tree level.

There are three diagrams at
one loop that contribute to the mass term of such scalars.
The first is an annulus diagram with the two scalar vertex
operators inserted on the same boundary.
The second is a Moebius diagram
with the two scalar vertex
operators inserted on the only boundary of the surface.
Both of these are drawn in figure \ref{fig1}.
The third diagram is shown
in figure \ref{fig2} and involves an annulus with the two vertex operators
 inserted on opposite
boundaries\footnote{This diagram
was not considered in the early analysis of \cite{pop}.}.

\ifPDFLATEX

\begin{figure}
\begin{center}
\includegraphics[width=15cm]{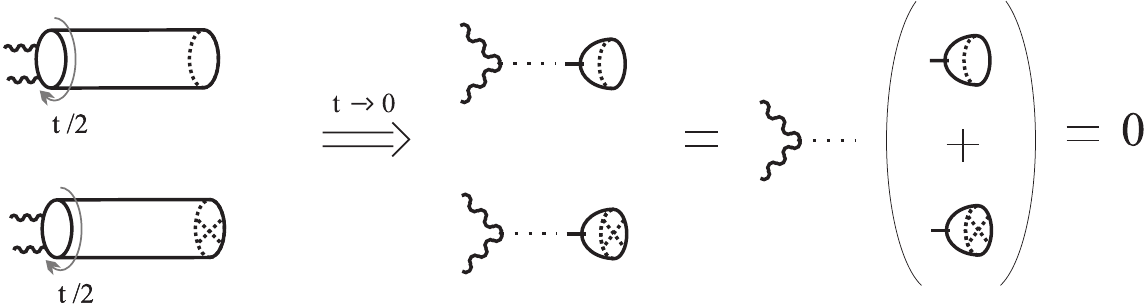}
\label{fig1}
\caption{Annulus and Moebius diagrams with the two scalars inserted on the same boundary and their UV factorization.}
\end{center}
\end{figure}

\begin{figure}
\begin{center}
\includegraphics[width=15cm]{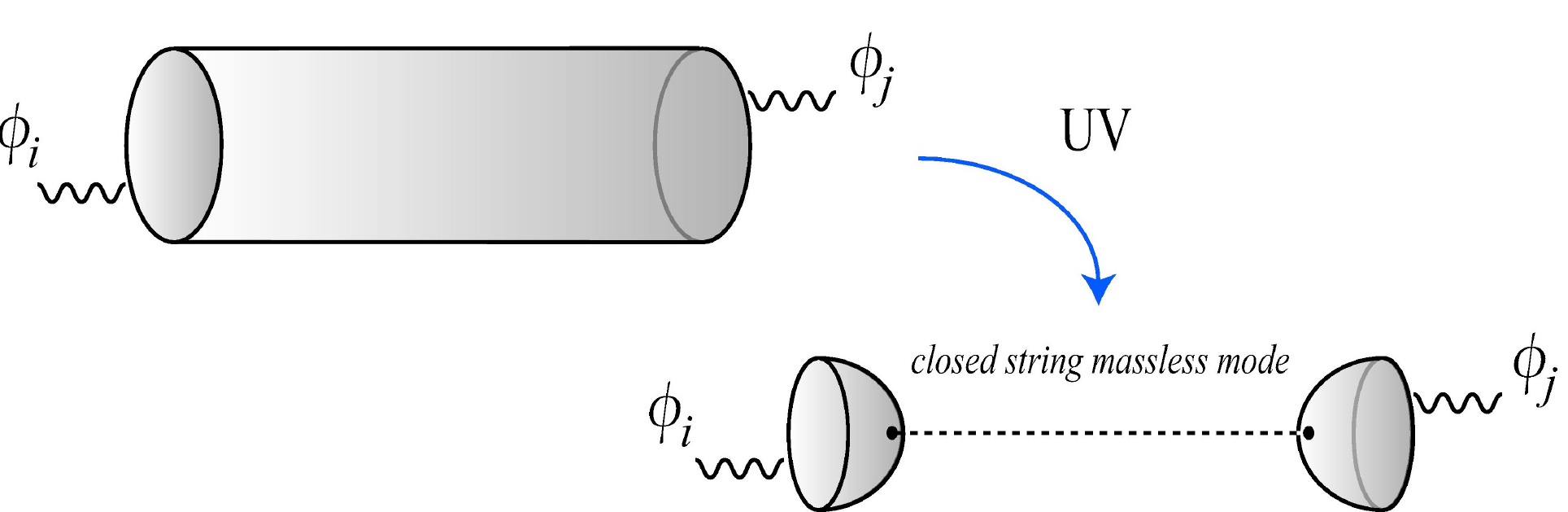}
\label{fig2}
\caption{Annulus  diagram with the two scalars inserted on opposite boundaries and its UV factorization.}
\end{center}
\end{figure}

\fi

All of these diagrams are
very similar in structure to the ones we must consider in order to calculate
the mass of anomalous U(1) gauge bosons in orientifolds \cite{akr}.
In such diagrams, the two
vertex operators give a kinematical piece that is ${\cal O}(p^2)$.
Therefore, to obtain a
contribution to the  mass
that is ${\cal O}(1)$ as
the momentum is small ($p^2\to 0$), we must obtain an $1/p^2$
pole from the
integration over the moduli of the surface.
There are two corners  such divergent terms can appear. In the
open-string IR channel,
this divergence is logarithmic
at best (or finite). The only source
of the pole is in the UV, and it is a contact term.
It can be obtained by
going to the transverse
closed string channel and
then looking at a massless divergence. At that limit the diagrams
factorize as shown in figures \ref{fig1} and \ref{fig2}.

For the two diagrams
of figure \ref{fig1} the residue of the
$1/p^2$ pole  is given by a product of a tree-level three-point coupling
that couples a scalar and
its conjugate to a massless
closed string mode, and the sum of the disk level tadpoles.
Therefore, if tadpoles
cancel at tree level this contribution is identically zero.

On the other hand,
in the diagram of figure \ref{fig2} the
residue of the $1/p^2$ pole
is a product of two disk two
point functions, each of them mixing the charged
opens-string scalar to a
closed string massless state.
However, if the U(1) symmetry
is intact at tree level such two-point mixing terms are identically zero.

Therefore, there are no one-loop
corrections to FI terms in orientifold
vacua under the conditions spelled out earlier.
It should be noted that as the
arguments above assume the background of CFT, they are not automatically
 applicable to vacua that contain
RR fluxes\footnote{They seem though to be valid perturbatively
in the RR field insertions.}.

Once the one loop correction
is zero no further perturbative or non-perturbative corrections are expected.

Returning to our vacuum, we
deduce that the FI are zero at the Gepner point, but
they may be non-zero if we move in some directions of the
closed string moduli space. The moduli along these directions are in the same
 chiral multiplets as the axions that cancel the anomalies of the relevant U(1)
symmetries.  Therefore it is necessary to not move in these directions.

\section{Instanton corrections}

 As we have already seen,
there is a remaining problem towards the phenomenological
  viability of the string vacuum under study, namely
 that there is no source for the masses of the $\cu,\cd$ quarks
 and the $\ch$ leptons: a whole family is so far massless.

 The missing couplings violate
the charge conservation of the two anomalous U(1) symmetries.
  We expect that instanton effects (both gauge instantons and stringy
 instantons) must
 non-perturbatively violate these symmetries. This is therefore
 a source for the missing couplings.

 Spacetime instantons in  string
theory have been analyzed for the first time after
  the advent of non-perturbative
duality symmetries, (see \cite{i-review} for a review).
 Their study has obtained a
boost recently \cite{stringinst} as it became obvious that
 they are crucial for several phenomenological questions in orientifold vacua,
 from generating neutrino masses to
Yukawa couplings to triggering supersymmetry breaking.

 In our case to generate the
relevant terms needed we need two kinds of instantons:
  one that
violates ($U(1)_c,U(1)_d$) charges  by (-1,1) units that we will call $I$
 and a conjugate one $I^*$ that
violates charges
by (1,-1) units\footnote{This cannot be the anti-instanton of I, as
 supersymmetry forbids the generation
of superpotential couplings in that case.}.
 In the case they may
generate the following non-perturbative superpotential up to cubic order
(further details are beyond the scope of the paper)
 \be
 W_I^{np}=  QUL+{Q{\cal D}\bar H}+LLE+L\bar H \sp W_{I^*I^*}^{np}=NN+NNR
 \label{9}\ee
\be
W_{I^*}^{np}={Q{\cal U}K}+EKK+K\bar H N+{N+NR+ NRR}
\label{10}\ee

It is important to arrange that the instantons do
not violate the $Z_2$ discrete symmetry,
 in which case the surviving non-perturbative superpotential reads
\be
 W^{np}={Q{\cal D}\bar H}+{Q{\cal U}H}+E\ch H+\ch N\bar H+NN+NR
\label{11}\ee
and as expected provides
Yukawa couplings for $\cu,\cd$ quarks, the $\ch$ lepton and the neutrinos.

In principle one can search for boundary states with the required number of zero-modes
to produce the required stringy instantons. However, there are several complicating issues that
have to be dealt with, such as the fact that we are not in the exact RCFT point (which may lead to
differences in the number of non-chiral zero-modes), the postulated $Z_2$ symmetry, the possibility
that undesired zero-modes may be lifted by fluxes, which we cannot take into account in the present
formalism, the fact that tree-level couplings between physical fields and zero-modes are needed,
plus the fact that not all boundary states present in the continuum
may be accessible within the context of RCFT.
For this reason a negative result would not be conclusive anyway, and we will not investigate this further
in the present paper, but take as our working hypothesis that the required instanton corrections exist.

\section{Neutrino masses}

An important ingredient in any realisation of the Standard Model is whether
 neutrino masses near what is measured today are possible.
A favourite mechanism for generating such neutrino masses is the see-saw
 mechanism and as
we will see a version of this mechanism is possible in our vacuum.

We will recollect here the
superpotential that is
relevant for neutrino masses from (\ref{w2}), (\ref{w3}),
(\ref{w4}) and (\ref{11}).
It includes both renormalizable and non-renormalizable contributions as
well as non-perturbative effects.
\be
W_{\n}={RR +LN\overline{H}}+
{\ch\bar H R+\ch\ch\bar H^2}+{L\bar HR+\ch\bar H N+NN+NR}
\label{12}\ee

\begin{table}
\begin{center}
{\large
 \begin{tabular}[t]{|c|c|c|c|c|c|c|}\trule

 & { $L^1$} & {$L^2$} & { $\ch$} & { $N$} & { $R^1$}& { $R^2$}\\ \trule

 $L^1$&0 & 0& 0& $v$ & $v e^{-S}$ &  $v e^{-S}$ \\\trule

 $L^2$ & 0& 0& 0& $v$ &  $v e^{-S}$ & $v e^{-S}$ \\\trule

 $\ch$ &0 &0 &$ v^2\over M_s$ & $v e^{-S}$ & $v$ & $v$  \\\trule

  $N$ & $v$ & $v$ &  $v e^{-S}$ & $M_s e^{-2S}$&  $M_s e^{-S}$& $M_s e^{-S}$ \\\trule

  $R^1$   &  $v e^{-S}$ & $v e^{-S}$ & $v$ & $M_s e^{-S}$ & $M_s$ & $M_s$\\\trule

    $R^2$  & $v e^{-S}$ & $v e^{-S}$ & $v$ &  $M_s e^{-S}$& $M_s$ & $M_s$ \\\trule
\end{tabular}}
\end{center}
\caption{Order of magnitude estimates of the  Neutrino mass matrix elements.
$v$ stands for the Higgs
vev, $M_s$ is the
string scale, and $e^{-S}$ stands for an instanton contribution.}
\label{tab4}\end{table}

The order of magnitude of the contributions of each term in the superpotential
 to the neutrino mass matrix is summarized in
table \ref{tab4}.
In this table the Higgs vev is labeled as v, the string scale $M_s$ is expected
to be near the unification scale, and the instanton factors
are sketchily labeled $e^{-S}$ and they can be small.

It is a straightforward
numerical exercise to verify that a matrix such as that in table
\ref{tab4} can reproduce
neutrino masses as suggested by experiment with
O(1) coefficients\footnote{We thank P.
Anastasopoulos for doing this calculation.}.

\section{Gauge couplings and unification}

In orientifold models the  hypercharge is given by\footnote{We neglect here the
 possibility that traceless generators appear in the hypercharge.
This happens many times, \cite{adks}, but is not
relevant for the vacua studied here.}
\be
Y=\sum_i k_i ~Q_i
\ee
where $Q_i$ are the overall U(1)
generators of $U(N_i)$ groups coming from complex brane stacks.
{}From this we can determine,
\cite{akt,d-review-2}, the hypercharge
coupling constant in the standard field theory normalization
as follows
\be
{1\over g_Y^2}=\sum_i~{2N_i\over g_i^2}
\label{13}
\ee
where $g_i$ is the gauge coupling of $i$-th stack, at the string scale.
These are determined at the tree-level by the string coupling and other
moduli, like volumes of
longitudinal dimensions as well as potential internal magnetic fields.
 At higher orders, they also
receive string threshold corrections.

For our vacuum with the hypercharge embedding (\ref{1})
we obtain
\be
{1\over g_Y^2}={1\over 6g_a^2}+{1\over 2g_c^2}+{1\over 2g_d^2}
\label{14}
\ee
from which we may compute the ${\rm sin}^2\theta_W$  at the string
scale
\be
\sin^2\theta_W\equiv{g_Y^2\over g_b^2+g_Y^2}=
{1\over 1+{g_b^2\over 6g_a^2}+{g_b^2\over 2g_c^2}+{g_b^2\over 2g_d^2}}\, ,
\label{15}
\ee
We have neglected  stringy
thresholds here, but they can be computed following \cite{ba}.

At the Gepner point and at
the string scale,  $g_a={g_b\over \sqrt{2}}=g_c=g_d$.
The extra factor for $g_b$ appears
because the $b$ brane is a real brane and this changes the normalization of the gauge coupling.
Also (\ref{15}) gives
 \be
\sin^2\theta_W(M_s)={3\over 10}
\label{16}\ee
This value differs
from the usual GUT value $3/8$ by  20\%,

As shown in appendix A, there is no scale
at which the weak (SU(2)) coupling constant can become twice the strong coupling
constant as is the case at the Gepner point.
This suggests that a correct fit to the SM gauge couplings is possible if in the
appropriate position in moduli space, this relation is modified appropriately.
The best case is that one moves to a point in moduli space where $g_b$ becomes equal to $g_a$.
In such a case if we assume for example $g_a=g_b=g_c=g_d$
then
\be
\sin^2\theta_W(M_s)={6\over 13}
\label{166}\ee
that differs from 3/8 by about 20\%.
In this case we show in appendix A that
 the standard unification ratio can be adjusted by
 lowering
the mass scale of non-chiral exotic multiplets  below the string scale.
Of course several other intermediate possibilities are also allowed.

\section{On supersymmetry breaking via gaugino condensation}

The hidden sector gauge group, SU(2) has a coupling that becomes
strong provided its chiral multiplets have masses close to the string scale.
This will drive gaugino condensation and   can break supersymmetry.

At the scale where the $SU(2)_h$ gauge group becomes strongly coupled the
corresponding gaugino condensate can trigger the supersymmetry breaking
\cite{Ferrara:1982qs}--\cite{Dine:1985rz}
(see \cite{Nilles:2008gq}--\cite{Quevedo:1996sv} for a review).
In particular the supersymmetry
breaking terms in the low energy effective action
have the form
$\frac{1}{M_{str}}  \int d^2 \theta  W^\alpha W_\alpha \Phi \Phi$,
where $W_\alpha$ is a
chiral superfield whose lower component is the gaugino $\lambda_\alpha$
and $\Phi$ is a matter
chiral superfield\footnote{Am alternative
mechanism of the supersymmetry breaking
via the gaugino condensate has been suggested in
\cite{Kaplan:1999ac}--\cite{Chacko:1999mi}
 in the framework of the brane world scenario.
In these models the
Standard Model gauge fields
are propagating in the bulk, while the matter is localized on the
brane. The value of the mass
terms in these models  depend on the size of the extra dimension.}.
 After the gaugino condensate
develops
a vacuum expectation value the
mass term of the form $\frac{<\lambda \lambda>}{M^2_{str}} $
can be generated.
The value of the gaugino condensate is related to the scale $\Lambda$
as $<\lambda \lambda> \sim \Lambda^3$
(an exact relation for the
case of $SU(2)$ gauge group can be found in \cite{Novikov:1983ee}).
{}From this  relation,
we must have  $\Lambda\sim 10^{11.7}$ GeV
in order to have a supersymmetry breaking scale of the correct magnitude.

To  estimate a scale where the hidden sector gauge group $SU(2)$
becomes strongly coupled we use the equation
\begin{equation}
\Lambda = M_{s} ~e^{ \frac{1}{2b_h {\tilde \alpha}(M_{s})}},
\end{equation}
where
\begin{equation}
b_h= 2 N_T + \frac{1}{2} N_X -6,
\end{equation}
and $N_T$ and $N_X$ are number of the chiral superfields $T$ and $X$
from the hidden sector which contribute to
the corresponding one loop beta-function.
We take
${\tilde
\alpha}^{-1}(M_{str.}) \sim 323.5$.
One can  consider different values for $b_h$.
 Let us first take
the case that no chiral
superfields contribute to $b_h$, ($N_T=N_X=0$) i.e., one
has only the contribution from the gauge bosons. One gets $\Lambda \sim
10^{4.2}$ GeV.
Another case is when one  $X$-field contributes
to $b_h$ ($N_T=0,N_X=1$). In this case one has
$\Lambda \sim 10^{3.2}$ GeV.
If there is a contribution from
more than one field $X$, the corresponding value of $\Lambda$
will lie below the scale $M_Z$.

To obtain a high enough value of the
gaugino condensation scale thresholds of KK states must be invoked.
A direct computation shows that if
the compactification scale is of
the order of $10^{15} GeV$, KK descendants of the SU(2) vector multiplet
will drive the SU(2) coupling strong at $\Lambda\sim 10^{11.7}$ GeV.

\section{Chiral symmetry breaking in the hidden sector}

The vacuum discussed here has a spectrum tabulated in table \ref{tab1}.
In particular, the hidden
sector SU(2)$_h$ has a chiral multiplet in the adjoint as well as 4 multiplets in the fundamental,
half of them carrying $Y={1\over 2}$ and the other half $Y=-{1\over 2}$.
Neglecting for the moment the SM interactions, there is an SU(4) chiral symmetry (the fundamental representation of SU(2)
is pseudoreal).

If we label the 4 SU(2)$_h$ doublet fermions by $X^I_{a,\a}$ where $a$ is the SU(2)$_h$ spinor index,
 $\a=1,2$ is the spin index and $I=1,2,3,4$ is a flavor index, then
$Y(X^{1,2}_{\a})={1\over 2}$, $Y(X^{3,4}_{\a})=-{1\over 2}$.
A gauge invariant order parameter for chiral symmetry breaking is
\be
Z^{IJ}=X^I_{a,\a}X^J_{b,\b}\epsilon^{\a\b}\e^{ab}\sp Z^{IJ}=-Z^{JI}
\ee
and its expectation value breaks chiral symmetry $SU(4)\to Sp(4)$ \cite{align}.

The alignment of the chiral condensate is however crucial concerning the (spontaneous)
 breaking of $U(1)_c$ and eventually electromagnetism.
As the limits on the photon mass are very stringent, this issue is of crucial importance
 in assessing the viability of this string vacuum.
The hypercharge of $Z^{12}$ is $Y=1$, that of $ Z^{34}$ is $Y=-1$ while the  other four $ Z^{IJ}$ have $Y=0$.

As in technicolor, the effective potential is generated by the exchange of the SM gauge
 bosons and it will prefer a direction where the $U(1)_c$
is unbroken, \cite{align}. As such directions exist, and are given by $Z^{12}=Z^{34}=0$,
 we conclude that for massless $X$ fields, $U(1)_{em}$
remains  unbroken. If we now move in moduli space, so that the X multiplets obtain an SU(4)
 invariant mass, we are guaranteed to remain at the same minimum
and $U(1)_{em}$ is still expected to remain unbroken.

So far our discussion above assumes the absence of supersymmetry. In the presence of unbroken supersymmetry,
the  potential for vacuum alignment due to the gauge interactions or masses is identically zero because of
supersymmetry. However, if eventually
supersymmetry is broken at a low scale then the potential discussed in the non-supersymmetric case
resurfaces and our earlier conclusions are valid.

\newpage

 \addcontentsline{toc}{section}{Acknowledgments}
\noindent {\bf Acknowledgements} \newline

~We ~are ~grateful ~to ~Pascal ~Anastasopoulos,
 ~Andres ~Collinucci, ~Maximilian ~Kreuzer,
~Hans-Peter ~Nilles, ~Michael ~Peskin ~and ~Stuart ~Raby, ~for
~valuable ~discussions ~and ~correspondence.

 This work was also partially supported by INTAS
grant, 03-51-6346, RTN contracts MRTN-CT-2004-005104 and
MRTN-CT-2004-503369, CNRS PICS \#~3747 and 4172,
 and by a European Union Excellence Grant,
MEXT-CT-2003-509661. The work of M.T. has been supported by Austrian
Research Funds, project P18679-N16 ``Nonperturbative effects in
string compactifications". The
work of A.N.S. has been performed as part of the program
FP 57 of the Foundation for Fundamental Research of Matter (FOM).

\newpage

\appendix
\renewcommand{\theequation}{\thesection.\arabic{equation}}
\addcontentsline{toc}{section}{Appendices}
\section*{APPENDIX}

\section{Analysis of the gauge couplings}

In this Appendix we give a brief
analysis of the renormalization group equations
for gauge coupling constants
(see also \cite{Blumenhagen:2003jy} for a similar discussion).
In particular we will show that
if the couplings at string scale are related by a relation similar to that of the Gepner point
point, $g_a={g_b\over \sqrt{2}}=g_c=g_d$,
(which in particular implies $\sin^2\theta_w={3\over 10}$) then
there is no way of fitting to the low energy coupling constants of the standard model.
In particular we will derive an upper bound for the weak coupling constant for this this to be possible.
We will then investigate another relation at the
string scale, namely   $g_a=g_b=g_c=g_d$, (which in particular implies $\sin^2\theta_w={6\over 13}$)
which as we show, fits the SM couplings, if some of the non-chiral exotics have masses below the string scale.
In general as we vary the appropriate closed string moduli, the couplings at the string scale will generically vary, and
the two relations we investigate here are two indicative cases.

We use the one-loop renormalization group equations
\begin{equation} \label{REGR1}
\frac{1}{\tilde \alpha_i(Q)}= \frac{1}{\tilde \alpha_i(\mu)}- 2 b_i
\log{ \frac{Q}{\mu}},
\end{equation}
where $\tilde \alpha_1= \frac{5}{3}\frac{g^2_Y}{16 \pi^2}$,
$\tilde \alpha_2= \frac{g^2_b}{16 \pi^2}$,
$\tilde \alpha_3= \frac{g^2_3}{16 \pi^2}$ and
$g^2_Y$, $g^2_b$ and $g^2_3$ are the coupling constants of
$U(1)_Y$, $SU(2)_b$ and $SU(3)$ gauge groups.
As we have mentioned before, we ignore the stringy threshold
corrections in the renormalization group equation (\ref{REGR1}).
The coefficients in the
renormalization groups equations without taking into account
 the contribution of the hidden sector fields are \cite{Jones:1981we}
\begin{equation} \label{bgn}
b_1 = \frac{4}{3}N_{Fam}+\frac{1}{10} N_{Higgs},
\quad b_2 = - \frac{22}{3}+\frac{4}{3}N_{Fam}+\frac{1}{6} N_{Higgs},
 \quad b_3 = - 11+\frac{4}{3}N_{Fam},
\end{equation}
for a case of a non-supersymmetric theory and
\begin{equation} \label{bgs}
b_1 = 2 N_{Fam}+\frac{3}{10} N_{Higgs}, \quad b_2 = -
6+2N_{Fam}+\frac{1}{2} N_{Higgs},
 \quad b_3 = - 9+2N_{Fam},
\end{equation}
for the supersymmetric case theories.
Here $N_{Fam}$ is a number of families of leptons and quarks and
$N_{Higgs}$ is a number of Higgs (super)fields.
Since we have three
families and two Higgs (super)fields the values of the coefficients $b_i$ are
\begin{equation}\label{bn}
b_1=\frac{21}{5}, \quad b_2 =-3, \quad b_3=-7,
\end{equation}
for energies below SUSY breaking scale and
\begin{equation}\label{bs}
b_1= \frac{33}{5}, \quad b_2=1, \quad b_3 = -3,
\end{equation}
for energies above SUSY breaking scale (that we take to be equal to $1$ TeV).
We assume that some of the non-chiral exotics  acquire masses
at an intermediate scale $M$
which is between the SUSY breaking scale and the string scale.
 Therefore these fields
contribute to the running of the coupling constants
at energies above the scale $M$.
The  corresponding contributions to the coefficients
$b_i$ are
\begin{equation}\label{dbh}
\Delta b_1= \frac{3}{10} N_{X}+ \frac{2}{5}N_{Y_s} + \frac{3}{5}N_{P_s} ,
\quad \Delta b_2 =0, \quad \Delta b_3= \frac{5}{2}N_{Y_s}.
\end{equation}
where $N_X$, $N_{Y_s}$ and $N_{P_s}$ are the numbers of superfields $X$, $Y_s$ and $P_s$
which get masses at the scale $M$.

We can now estimate the value of the scale $M$ by fitting the gauge couplings to the observable values.
Let us denote $3g_w^2/5g_Y^2= \frac{3}{5} ctg^2 \theta_W (M_s)\equiv \gamma $.
The renormalization group equation reads
\begin{equation}\label{RGSG}
\frac{1}{\tilde \alpha_1(M_{susy})}+ 2 b_1 \log{ \frac{M_{susy}}{M_{s}}}
+ 2 \Delta b_1 \log{ \frac{M}{M_{s}}}=
\gamma(\frac{1}{\tilde \alpha_2(M_{susy})}+
2 b_2 \log{ \frac{M_{susy}}{M_{s}}}),
\end{equation}
{}From the equation (\ref{RGSG}) we observe that not all possible values of $\gamma$
are allowed, since the value of $\Delta b_1 \log{\frac{M_{s}}{M} }$
must be positive\footnote{It is in principal possible that stringy thresholds can bypass this constraint.}.
The limiting value of $\gamma$ corresponds to the case
of the ``standard'' unification of coupling constants i.e.,
$\gamma=1$ and $\Delta b_1=0$. Therefore $\gamma$ must be less or equal to $1$.
On the other hand  $\Delta b_1 \log{\frac{M_{s}}{M} }$ can not be too large, since
 it will imply
that the value of the scale $M$ is very low. Estimating the lowest possible value
of $M$ to be around 1 TeV we get the lowest value of $\gamma$ to be $\sim 0.26$
(this corresponds to the maximal value of $\Delta b_1$).
Therefore we conclude that the value of $\gamma$ must be between $0.26$ and $1$.

Therefore we conclude that the case $g_a= \frac{g_b}{\sqrt 2}= g_c=g_d$ is excluded
since in this case $\gamma= \frac{7}{5}$.
On the other hand for the case $g_a=g_b=g_c=g_d$ is allowed since
 $\gamma = \frac{21}{30}$.
Let us consider this case in more detail. The renormalization group equation now reads
 \begin{equation}\label{RGS}
\frac{1}{\tilde \alpha_1(M_{susy})}+ 2 b_1 \log{ \frac{M_{susy}}{M_{s}}}
+ 2 \Delta b_1 \log{ \frac{M}{M_{s}}}=
\frac{21}{30}(\frac{1}{\tilde \alpha_2(M_{susy})}+
2 b_2 \log{ \frac{M_{susy}}{M_{s}}}),
\end{equation}
where the coefficients
$b_i$ and $\Delta b_i$ are
given by (\ref{bs}) and (\ref{dbh}) and the values of
$\tilde \alpha_1(M_{susy})$
and $\tilde \alpha_2(M_{susy})$ can be obtained from
(\ref{REGR1}), (\ref{bn}) and their values at
$M_Z$ ($\sim 10^{2}$ Gev)
scale (see for example  \cite{Mohapatra:1999vv}-- \cite{Kazakov:2000ra} )
\begin{equation}
\frac{5}{3}\frac{g_Y^2(M_Z)}{4 \pi}= 0.017,
\quad \frac{g^2_b(M_Z)}{4 \pi}= 0.034, \quad
\frac{g^2_3(M_Z)}{4 \pi}= 0.118.
\end{equation}
{}From the equation (\ref{RGS})
one obtains (we have taken $M_{s} \sim 10^{16}$GeV )
\begin{equation}
\Delta b_1 \log{\frac{M_{s}}{M} }= 49.14,
\end{equation}
 Obviously the value $\Delta b_1$ and therefore the value
 of the scale $M$ depends on how many and which superfields from the hidden
sector contribute to the running of the coupling constant $g^2_Y$
between scales $M$ and $M_{s}$.
For example let us consider the case when all non-chiral exotics
contribute to the running of the coupling constant, i.e,.
$N_{Y_s}=4, N_X=4, N_{P_s}=2$. This  gives $\Delta b_1 = 4$, therefore
$\log{\frac{M_{s}}{M} }=12.3 $ and $M \sim 4.5 \times 10^{10}$ GeV.
Let us note that this case
will also change  the running of the strong coupling constant
comparing to the usual MSSM
because of $Y_s$ field.
Another possible
 case is  when fields $Y_s$ obtain their masses at the string scale, i.e.,
$N_{Y_s}=0, N_X=4, N_{P_s}=2$.
 One has $\Delta
b_1 = 2.4$, $\log{\frac{M_{s}}{M} }=20.5$ and $M \sim 1.25 \times
10^{7}$ GeV.
Another example is $N_{Y_s}=1, N_X=1, N_{P_s}=2$.
In this case one has $\Delta b_1=1.9$
and $M \sim 5.8 \times 10^{4}$ GeV.

Therefore
one can conclude that if some of the hidden sector fields obtain their masses
at an intermediate scale $M$ which is
between SUSY breaking scale and the string scale,
one can have a correct
fitting of gauge coupling constants at the string scale, which is
compatible with their low energy values.

\section{Minimisation of the Higgs potential}

 As it was explained
in the Section 5, the bosonic
component of the linear combination $H^1+H^2$ is expected to develop
a vacuum expectation value and will be therefore  identified with the Higgs field $H_u$.
Because of the presence of  singlets $R$
the Higgs potential is different from that of the MSSM
and we analyze its minimization here.

Ignoring the terms which come from fourth order terms in the superpotential
(like $HH {\overline H} {\overline H}$)
 the relevant part of
 the potential has the form
\begin{eqnarray}
V&=&m_1^2 H H^\dagger + m_2^2 {\overline H} {\overline H}^\dagger +
m_3^2 (H {\overline H}-{\overline H}^{ \dagger} {H}^\dagger)+
\frac{g^2}{8}{(H {H}^\dagger - {\overline H} {\overline
H}^\dagger)}^2
\\ \nonumber &&
+\frac{g^2}{2}(H {\overline H}^\dagger)({\overline H} H^\dagger)
+\eta^2 (H {\overline H})({\overline H}^\dagger H^{ \dagger})
+\frac{g_Y^2}{8} {(\xi_Y+ H H^\dagger - {\overline H} {\overline
H}^\dagger)}^2,
\end{eqnarray}
where the term proportional to the parameter $\eta$ comes from the
terms of the type $H {\overline H} R$ in the superpotential.
 Let us further take an ansatz for the Higgs fields as
\begin{equation}
H_1 = v_u, \quad {\overline H}_2 = v_d.
\end{equation}
The extremization conditions are
\begin{equation}
(m_1^2+ \frac{\xi g_Y^2}{4})  v_u - m_3^2 v_d + \frac{g^2 +
g_Y^2}{4}(v_u^2 - v_d^2)v_u + \eta v_u v^2_d =0,
\end{equation}
\begin{equation}
(m_2^2- \frac{\xi g_Y^2}{4})
 v_d - m_3^2 v_u -\frac{g^2 + g_Y^2}{4}(v_u^2 - v_d^2)v_d
+ \eta v_u^2 v_d =0.
\end{equation}
Introducing the parametrization $v_u = v \cos \beta$ and $v_d = v
\sin \beta$ we can solve the last two equations
\begin{equation} \label{solu1}
v^2 = - 4\frac{{\tilde m}_1^2 - {\tilde m}_2^2 \tan^2 \beta}{(g^2
+ g_Y^2)(1-\tan^2 \beta) + 8 \eta^2 \sin^2 \beta}
\end{equation}
\begin{equation} \label{solu2}
\sin 2 \beta = \frac{2 m_3^2}{{\tilde m}_1^2 + {\tilde m}_2^2 + \eta^2}
\end{equation}
where we have denoted ${\tilde m}_1^2= m_1^2+ \frac{\xi g_Y^2}{4}$
and ${\tilde m}_2^2= m_2^2- \frac{\xi g_Y^2}{4}$. The gauge symmetry
breaking condition (i.e., the conditions
 that the solution
(\ref{solu1}-- \ref{solu2} is the
minimum) are ${\tilde m}_1^2 {\tilde m}_2^2 < m_3^4$ and
$2 m_3^2 < {\tilde m}_1^2 + {\tilde m}_2^2 + \eta^2$

\end{document}